\documentclass[namedreferences]{SolarPhysics}
\usepackage[optionalrh]{spr-sola-addons} 
\usepackage{graphicx}        
\usepackage{color}           
\usepackage{url}             

\newcommand{\aap}{    {\it Astron. Astrophys.}}
\newcommand{\aaps}{   {\it Astron. Astrophys. Suppl.}}

\newcommand{\apj}{    {\it Astrophys. J.}}

\newcommand{\solphys}{{\it Solar Phys.}}
\newcommand{\sovast}{ {\it Sov.  Astron.}} 
 
\newcommand{\araa}{	{\it Ann. Rev. Astron. Astrophys.}}

\begin{document}

\begin{article}

\begin{opening}

\title{Allen Telescope Array Multi-Frequency Observations of the Sun}

\author{P.~\surname{Saint-Hilaire}$^{1}$\sep
        G.J.~\surname{Hurford}$^{1}$\sep
        G.~\surname{Keating}$^{2}$\sep 
        G.C.~\surname{Bower}$^{2}$\sep
	C.~\surname{Gutierrez-Kraybill}$^{2}$
       }
\runningauthor{P. Saint-Hilaire {\it et al.}}
\runningtitle{ATA solar observations}

   \institute{$^{1}$ Space Sciences Laboratory, University of California, Berkeley, CA 94720, USA
                     email: \url{shilaire@ssl.berkeley.edu} email: \url{ghurford@ssl.berkeley.edu}\\ 
              $^{2}$ Radio Astronomy Laboratory and Department of Astronomy, University of California, Berkeley, CA 94720, USA
                    email: \url{gkeating@astro.berkeley.edu} email: \url{gbower@berkeley.edu} \\
             }

\begin{abstract}

	We present the first observations of the Sun with the Allen Telescope Array (ATA).
	We used up to six frequencies, from 1.43 to 6 GHz, and baselines from 6 to 300 m.
	To our knowledge, these are the first simultaneous multi-frequency full-Sun maps obtained at microwave frequencies without mosaicing.
	The observations took place when the Sun was relatively quiet, although at least one active region was present each time.
	We present multi-frequency flux budgets for each sources on the Sun.
	Outside of active regions, assuming optically thin bremsstrahlung (free--free) coronal emission on top of an optically thick $\approx$10\,000 K chromosphere,
	the multi-frequency information can be condensed into a single, frequency-independent, ``coronal bremsstrahlung contribution function'' $[EM/\sqrt{T}]$ map.
	This technique allows the separation of the physics of emission as well as a measurement of the density structure of the corona.
	Deviations from this simple relationship usually indicate the presence of an additional gyroresonance--emission component, as is typical in active regions.
	
\end{abstract}
\keywords{Corona, Radio Emission; Radio Emission,  Quiet; Radio Emission,  Active Regions }
\end{opening}

\section{Introduction}

	The non-flaring Sun has been studied at radio wavelengths for several decades.
	(For a recent review, see {\it e.g.} \opencite{Shibasaki2011}).
	Most of the quiescent solar microwave emission is due to free--free bremsstrahlung (FF) emission from the lower corona and upper chromosphere.
	At microwave frequencies, the quiet--Sun emission can be modeled as an optically thin $\approx$1 MK corona on top of an optically thick $\approx$10\,000 K chromosphere ({\it e.g.} \opencite{Zirin1991}; but see \opencite{LandiChiuderiDrago2003} and \citeyear{LandiChiuderiDrago2008} for a nuance).
	In active regions (AR), an additional broadband gyroresonance component, typically peaking around 2800 GHz (10.7 cm), is present when the coronal magnetic field is strong enough \cite{Zlotnik1968a,Zlotnik1968b,WhiteKundu1997}.
	During flares, additional thermal bremsstrahlung and gyrosynchroton components occur ({\it e.g.} \opencite{Dulk1985}).

	Recent efforts to understand microwave emission from the Sun's upper atmosphere have concentrated on either the spectral aspect, with multi-frequency measurements at Sun center \cite{Zirin1991,Borovik1992,LandiChiuderiDrago2003,LandiChiuderiDrago2008}, 
	or by studying its spatial profile at a few frequencies ({\it e.g.} \opencite{Selhorst2003}, \opencite{Selhorst2005}, \opencite{Selhorst2010}, \opencite{Krissinel2005} for recent work on the topic).
	However the two approaches have not yet been reconciled into producing a density model that satisfactorily explains the equatorial limb profile in the whole microwave band, let alone the polar one.
	
	The 10.7 cm spatially integrated solar radio flux is known to be well correlated with the sunspot number ({\it e.g.} \opencite{Tapping1994}).
	Qualitatively, the reason is simple:  active regions tend to have both enhanced densities (leading to increased free--free emission), and an additional gyroresonance (GR) component. 
	But, to our knowledge, a systematic observational study of the fraction of the full-Sun 10.7 cm flux that is due to GR and that is due to FF emission remains to be done.
	
	Section~\ref{sect:ATA} of this article introduces the Allen Telescope Array in the context of solar observations. 
	Section~\ref{sect:obs} presents the three sets of ATA solar observations obtained so far, including raw brightness temperature $[T_{\textrm{B}}]$ maps and equatorial/polar profiles.
	For comparison, cross-calibration with other instruments or semi-empirical models is also presented.
	In Section~\ref{sect:budget}, a multi-frequency flux budget for each observation is presented, and the fraction of the solar 10.7 cm flux that is due to gyroresonance emission is determined.
	In Section~\ref{sect:CTBC}, we derive ``Coronal Thermal Bremsstrahlung Contribution'' (CTBC) maps and profiles from observations, a format which facilitates cross-frequency comparisons, 
	and allows for the quick discrimination between regions on the Sun where only optically thin free--free emission is relevant, and regions where additional physics ({\it e.g.} optical thickness, gyroresonance) is present.
	The results are summarized in Section~\ref{sect:ccl}.

\section{Solar observations with the Allen Telescope Array}\label{sect:ATA}

	The Allen Telescope Array \cite{Welch2009} is a radio interferometer near Hat Creek, California, USA.
	It consists of 42 6.1-m antennas located in an optimized ``pseudo-random'' configuration within a 300-m diameter.
	Two digital correlators can be tuned to analyze two 100-MHz wide bands (1024 channels each), independently chosen between 0.5 GHz and 11.2 GHz. Each of these observing frequencies can be changed (under computer control) in a few seconds. 
	Through careful design, the interferometer system is phase-stable and is fully calibrated (in amplitude and in phase) with respect to standard cosmic sources.
	The gains and signal levels of the receivers and digital back-end can be reconfigured under computer control to handle expected solar signal levels. 
	Temporal resolution as high as one second is currently available from the correlators. 
	The correlator output can be transformed into images through standard techniques using a well-developed
	data analysis package ({\sf Miriad}:\url{http://bima.astro.umd.edu/miriad/} \opencite{Sault1995}) for semi-automated calibration and mapping with pipeline software for flagging, calibration, and imaging specific to the ATA \cite{Keating2009}.

	The ATA is well-suited for imaging/spectroscopy of large--scale, slowly varying solar phenomena. 
	It is the only instrument currently available that can provide instantaneous two-dimensional images of the full Sun at frequencies near 2800 MHz. 
	The 6.1-m antennas provide a field of view that scales with frequency ($HPBW \approx \frac{2.8^{\circ}}{\nu (Ghz)}$; {\it viz.} full-Sun field of view when $\nu \lesssim $6 GHz).
	The interferometer's angular resolution also scales with frequency ($HPBW \approx \frac{4.5'}{\nu (Ghz)}$).
	Even though this is insufficient to resolve the internal structure of active regions or burst sources, it is well-suited to separating the various spatial contributions from across the disk. Furthermore, the compact array includes (projected) baselines as short as $\approx$6 m, corresponding to fringe spacings as large as $\approx$one degree. 
	Therefore in addition to imaging, the system can provide information on large--scale structures ({\it e.g.} coronal holes) with size scales of 10--20 arcminutes as well as an independent measurement of the total solar flux.

	Although the ATA has dual linearly polarized feeds, for this application the correlator combines these to measure circular polarization, thereby providing an additional diagnostic of solar sources. 
	Investigations of compact radio sources \cite{Law2010} indicate that the accuracy of Stokes $V$ maps is $\approx$1\%.

\section{Summary of Observations}\label{sect:obs}

		\begin{figure}[p]
		\centering
		\includegraphics[width=10cm]{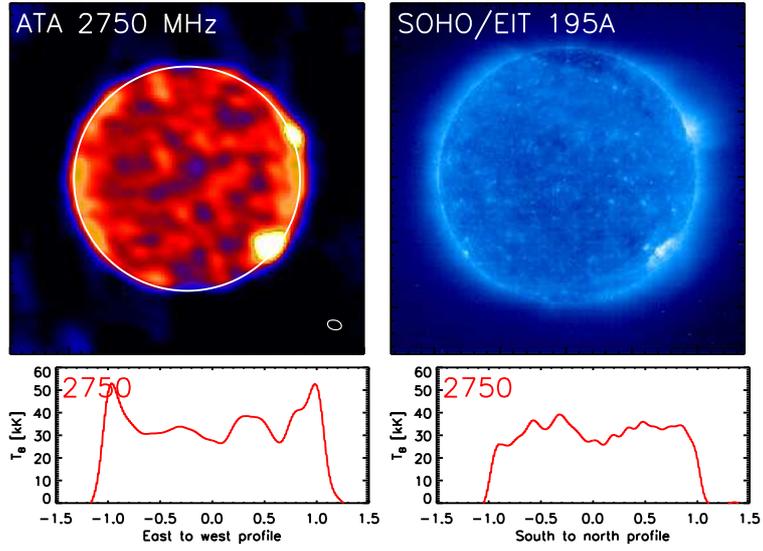}
		\caption{
			1-Oct-2009 observation (two-minute integration): brightness temperature $T_{\textrm{B}}$ maps and profiles.
			Top left: $T_{\textrm{B}}$ maps, with photospheric limb and FWHM beam sizes in the lower--right corner.
			Top right: Corresponding SOHO/EIT EUV image.
			Bottom left: equatorial $T_{\textrm{B}}$ profile.
			Bottom right: polar $T_{\textrm{B}}$ profile.
			All maps in this work have solar North oriented ``up''.
		}
		\label{fig:img:Oct01:4a}
		\end{figure}

		\begin{figure}[p]
		\centering
		\includegraphics[width=12cm]{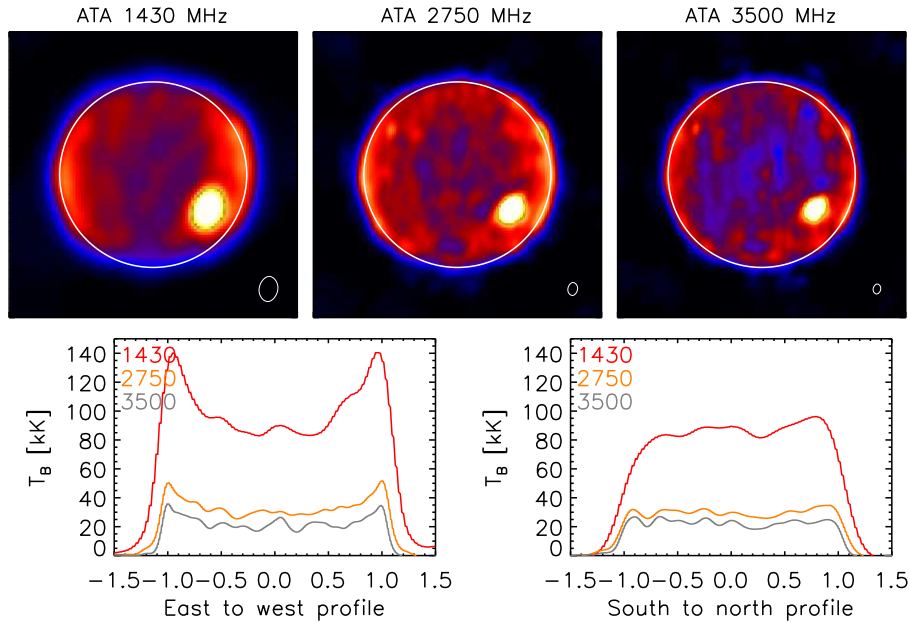}
		\caption{
			29-Jan-2010. Brightness temperature maps (with photospheric limb contour and FWHM beam sizes in the lower--right corners) and profiles.
		}
		\label{fig:img:Jan29:4a}
		\end{figure}

		\begin{figure}[t]
		\centering
		\includegraphics[width=12cm]{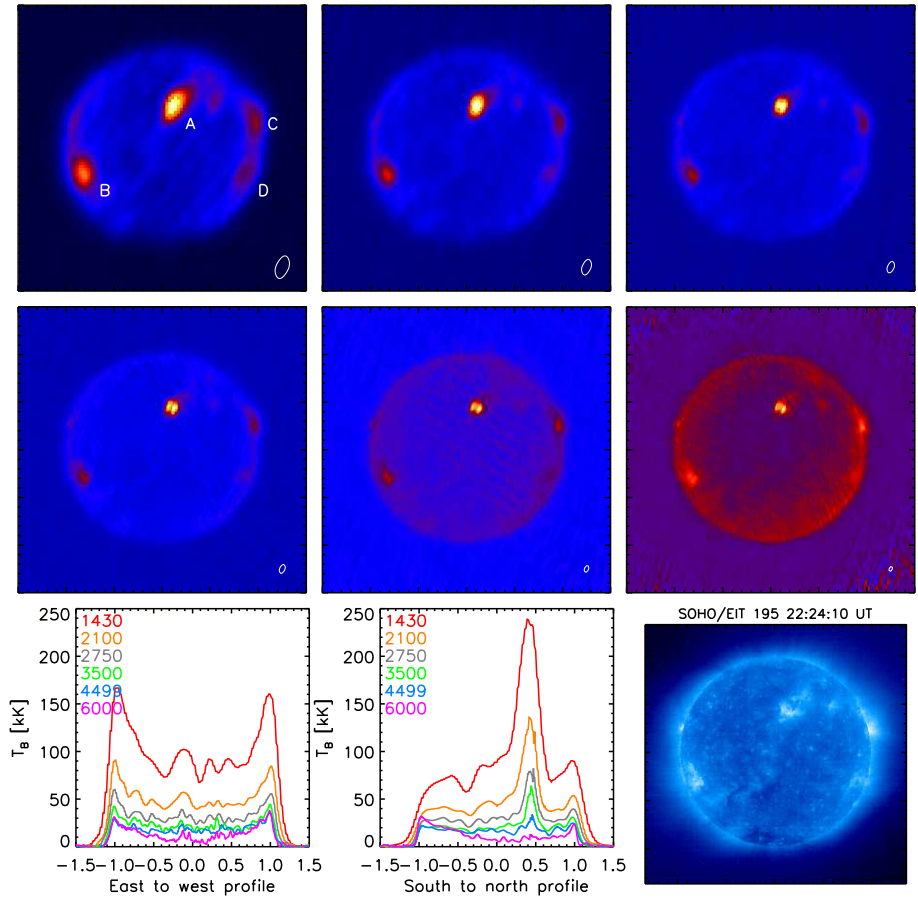}
		\caption{
			21-Mar-2010: Brightness temperature $T_{\textrm{B}}$ maps at 1430, 2100, 2750, 3500, 4500, and 6000 MHz,
			along with HPBW beamsizes in the lower--right corner.
			Bottom row: $T_{\textrm{B}}$ equatorial and polar profiles (through Sun center), and the corresponding SOHO/EIT image.
			The profiles (polar in particular) display an important contribution by the active region.
		}
		\label{fig:img:Mar21:4a}
		\end{figure}

		\begin{figure}[t]
		\centering
		\includegraphics[width=12cm]{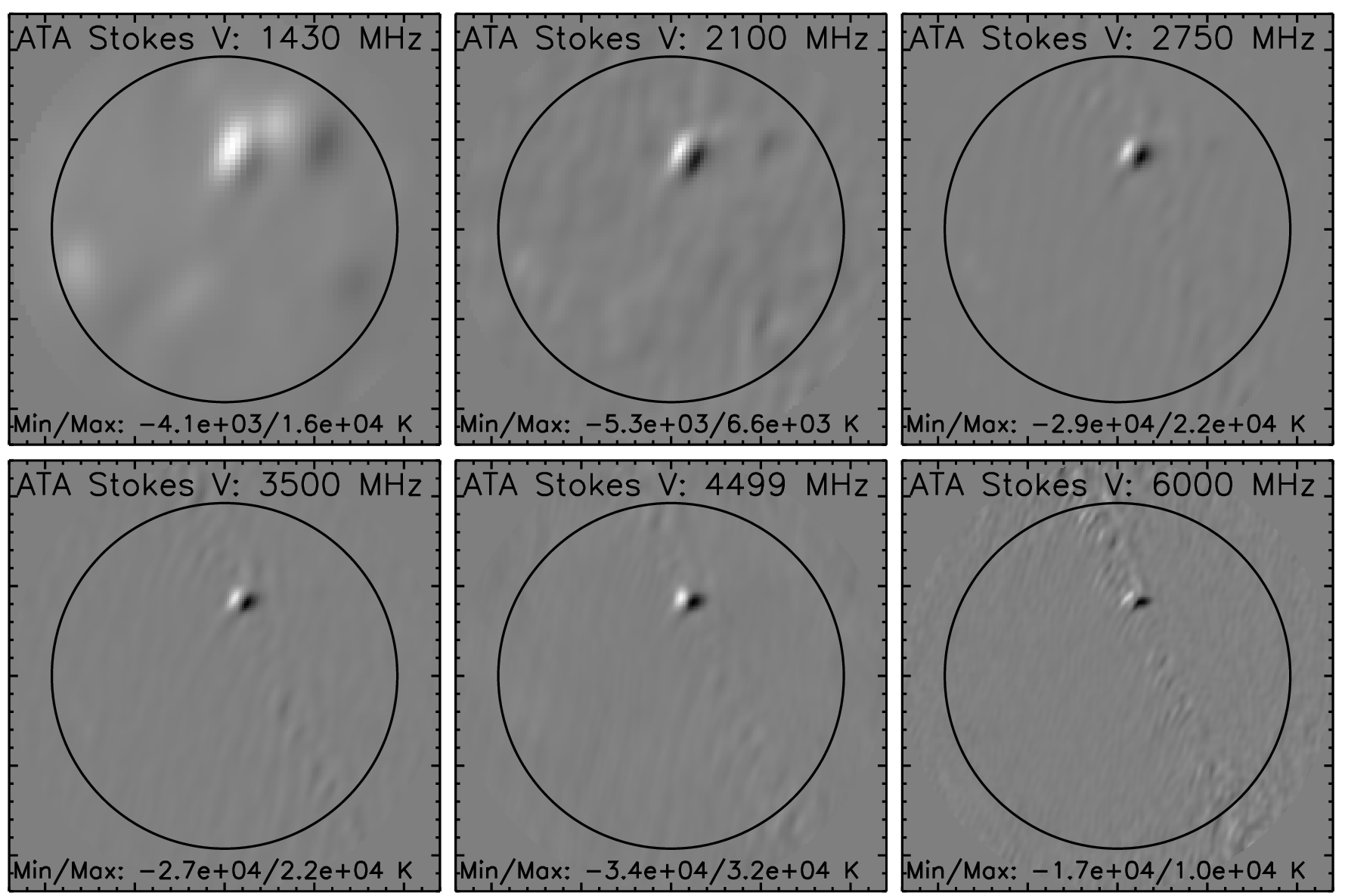}
		\caption{
			Stokes $V$ maps for the 21-Mar-2010 observations.
			The non-negligible level of polarization (varying with frequency, peaking at 11\% at 4.5 GHz) in the AR is a good indicator that magnetobremsstrahlung emission is also present.
			As there is some spatial smearing happening (ATA HPBW at 6 GHz is $\approx$45'', while the typical magnetic size scale in an AR is $\approx$10'', \protect\opencite{Lee1993a} ), the degree of polarization may be underestimated.
		}
		\label{fig:img:Mar21:4a:V}
		\end{figure}

	Three sets of observations were taken:
	\begin{enumerate}
		\item {\bf 1-October-2009} (Figure \ref{fig:img:Oct01:4a}):
			In this first test, the Sun was observed for two minutes around 21:56:36 UT, at 2750 MHz only.
			No flare activity was detected during that time (the GOES X-ray digitization was bottoming out at the lowest observable A0.3 level).
			Two small active regions were present, both near the western limb of the Sun.
		\item {\bf 29-January-2010} (Figure \ref{fig:img:Jan29:4a}):
			The Sun was observed with 31 antennas at 1430, 2750, and 3500 MHz, for $\approx$15 minutes, from 21:44 to 22:01 UT.
			The GOES X-ray level was low and stable between the A2 and A3 level (no flares). 
			Radio Solar Telescope Network (RSTN) lightcurves were flat.
			The only feature is a single on-disk active region.
		\item {\bf 21-March-2010} (Figure~\ref{fig:img:Mar21:4a}):
			The Sun was observed at six frequencies centered at 1430, 2100, 2750, 3500, 4500, and 6000 MHz,
			for eight hours beginning at 16:30 UT,
			in 1-minute intervals, alternating between frequency pairs.
			There were several active regions: a main on-disk AR (``A'' in Figure~\ref{fig:img:Mar21:4a}), and several smaller ones near both equatorial limbs (``B'' to ``D'' in Figure~\ref{fig:img:Mar21:4a}).
			The Sun was slightly active: the GOES X-ray baseline was A6--A8, and four small B-class flares (B1.0 to B2.2) occurred during the time the data were accumulated.
			These have negligible influence on our results.
			We have also plotted the Stokes $V$ maps for this observations (Figure~\ref{fig:img:Mar21:4a:V}), which will be discussed in Section~\ref{sect:budget}.
	\end{enumerate}

	Each ATA map was CLEANed \cite{Hogbom1974} around bright active region features, and the rest of the image ($\approx$quiet Sun) was then treated with a Maximum Entropy Method (MEM).
	Image dynamic range (ratio of brightest pixel to rms noise of image) for the 21-Mar-2010 observations is about 700 at 1430 MHz, and 250 at 6000 MHz.

	We discuss the accuracy of the flux calibration in the next sub-sections, by comparing total map fluxes to that given by other (spatially integrated) instruments (Section~\ref{sect:totmapflux}), 
	and by comparing the brightness temperature at Sun center with other published results (Section~\ref{sect:tbsuncenter}).
	We then say a few words about the brightness temperature profiles and Stokes $V/I$ maps.

	\subsection{Total Map Fluxes:} \label{sect:totmapflux}
	
		\begin{figure}[h]
		\centering
		\includegraphics[height=11cm]{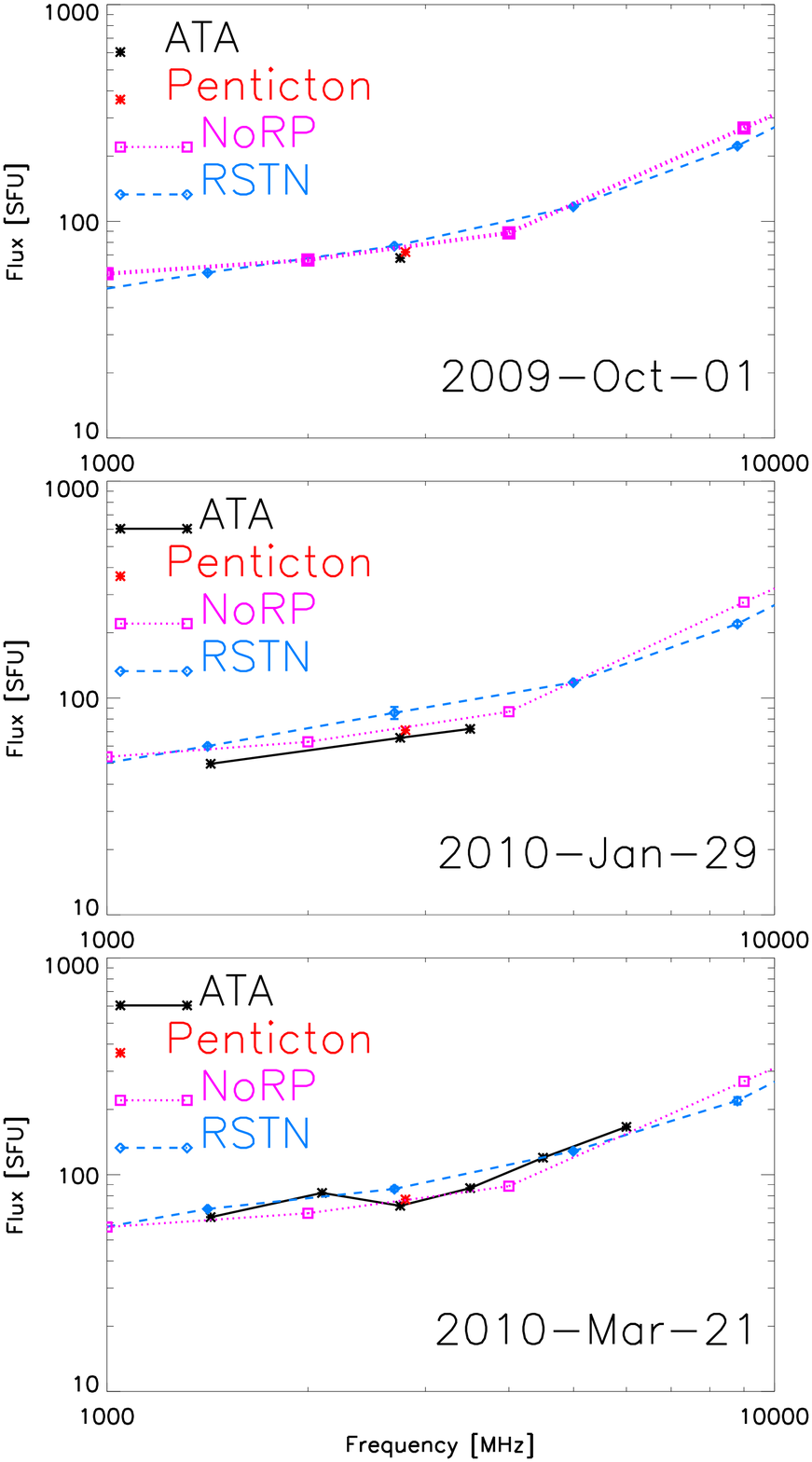}
		\caption{
			Full-map fluxes from ATA and comparisons with other observatories.
			Black: ATA
			Red: Penticton, adjusted noon value.
			Blue: Radio Solar Telescope Network (RSTN) spectrum, averaged between the Palehua and Sagamore Hill observatories.
			Purple: Nobeyama Radio-polarimeter (NoRP) values.
		}
		\label{fig:sp:comparisons}
		\end{figure}

	Figure~\ref{fig:sp:comparisons} displays the total fluxes in each map at each frequency for all three periods of  observations, 
	with comparisons with other instruments: the 10.7 cm flux from Penticton (British Columbia), the Nobeyama Radiopolarimeter (NoRP; Japan), and two observatories that are part of the Radio Solar Telescope Network (RSTN; Palehua, Hawaii and Sagamore Hill, Massachussets), all adjusted to 1 AU.
	There is generally good agreement, typically within the 10--15\% level.

	\subsection{Center Pixel $T_{\textrm{B}}$:}\label{sect:tbsuncenter}
		
		\begin{figure}[h]
		\centering
		\includegraphics[width=12cm]{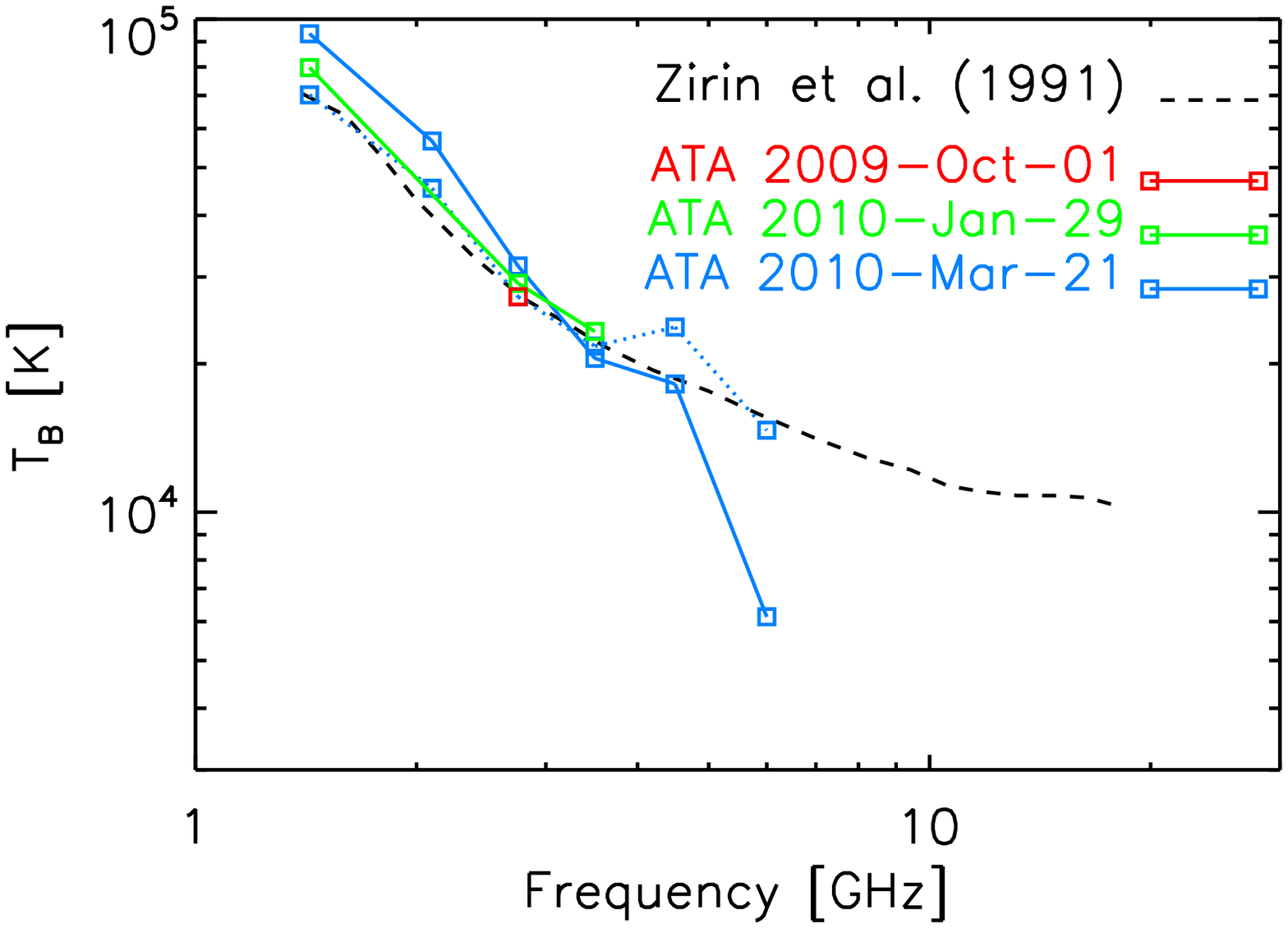}
		\caption{
			{\bf Center pixel $T_{\textrm{B}}$.}
			21-Mar-2010 observations (solid, blue) center $T_{\textrm{B}}$ are higher than expected at low frequencies because of contribution from a nearby AR.
			The dashed, blue line was taken instead for pixels about 500'' further south, corresponding to $\approx$2 HPBW (at 1430 MHz) away from main AR.
			The value of $T_{\textrm{B}}$ around Sun center is highly oscillatory (spatially) at high frequency, probably due to the lack of small baselines at these frequencies.
			Taking the modes of the maps lead to values in between the extremes shown (between the dashed and the solid blue lines).
		}
		\label{fig:sp:centerpixelTb}
		\end{figure}

	We compared the brightness temperature $T_{\textrm{B}}$ of the pixel nearest Sun center to the results published by \inlinecite{Zirin1991} (see Figure~\ref{fig:sp:centerpixelTb}).
	There is good agreement, except at 6 GHz, where we observe a high amount of on-disk spatial oscillations, and a low on-disk average brightness temperature, 
	which we attribute to a lack of short baselines at these frequencies.
	Hence, the values at high frequencies, particularly at 6 GHz, must be used with caution.

	\subsection{$T_{\textrm{B}}$ profiles:}\label{sect:tbprofiles}
	
	We point out the equatorial limb brightening observed at all frequencies and for all three observations, and the absence of such at the poles.
	This has already been observed before ({\it e.g.} \opencite{Christiansen1955}), at higher spatial accuracy.
	Despite this limitation, we note that ATA's multi-frequency capability allows us to study limb brightening in a new light (see introduction for references on the topic).
	We shall make a few comparisons with simple atmospheric models later in this work, but a thorough investigation is beyond the scope of the present article.
	Suffice it to say for now that the equatorial limb is thought to be brighter than the polar limb because the solar Equator comprises many closed loops, which tend to retain plasma,
	while polar regions, predominently composed of ``open field lines'' where the plasma can easily escape with the solar wind, are less dense.

	\subsection{Stokes $V/I$ Maps:}\label{sect:stokesV}
	
	Figure~\ref{fig:img:Mar21:4a:V}	shows a clear and consistent circular polarization signature accross all frequencies for the disk AR (source ``A''),
	and are consistent with SOHO/MDI magnetic maps of the underlying photosphere.
	The maximum Stokes $V/I$ occurs at 4.5 GHz, where it can reach 11\%.
	This value is probably a lower limit, as there is certainly some spatial smearing in effect: 
	the typical magnetic size scale in a sunspot is about 10'' \cite{Lee1993a},but the instrumental resolution is about 1' at 4.5 GHz. 
	Furthermore, solar rotation moves the source horizontally by $\approx$1' over the course of data accumulation.

	Free--free thermal bremsstrahlung emission in active regions is not expected to be able to reach such high levels of circular polarizations ({\it e.g.} Gelfreikh, \citeyear{GaryKeller2004}),
	and, in the absence of meaningful gyrosynchroton emission from radio bursts, we therefore conclude that it is mostly due to gyroresonance emission.

	\section{Flux Budget:} \label{sect:budget}

\begin{figure}[ht!]
		\centering
		\includegraphics[width=7.5cm]{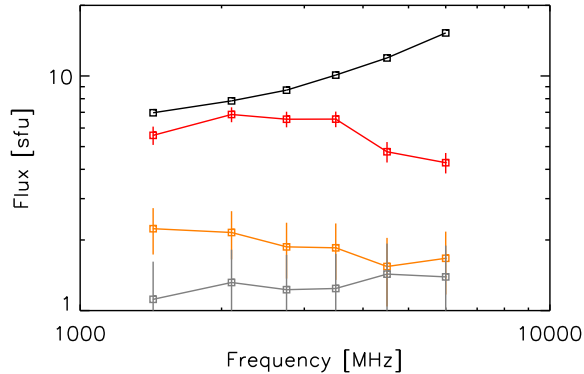}
		\caption{
			Spectra for the 21-Mar-2010 observations.
			All values were normalized using interpolated values of the solar fluxes observed by NoRP.
			From top to bottom:
			full-Sun value, divided by ten (black),
			background-subtracted main active region labelled ``A'' in Figure~\ref{fig:img:Mar21:4a} (red),
			background-subtracted region ``B'' (orange),
			and background-subtracted region ``C'' (gray).
			Source ``D'', being even weaker, is not displayed.
			Free--free emission is expected to have a flat flux spectrum (while it is optically thin), consistent with all sources besides ``A''.
			The ``bump'' at mid-frequencies for source ``A'' is a strong indicator of gyroresonance emission.
		}
		\label{fig:sp:Mar21}
		\end{figure}

	Tables~\ref{tab:budget:Oct01}, \ref{tab:budget:Jan29}, and \ref{tab:budget:Mar21} contain a flux budget for the most prominent sources in each of the three observations.
	The ``full-Sun'' flux was computed by adding together all pixel values within 1.3 $R_S$ of Sun center.
	Additionally, for the 21-Mar-2010 observations, source spectra have been plotted (Figure~\ref{fig:sp:Mar21}).
	As most of the features were spatially unresolved (at least at the lowest frequencies), we have plotted the fluxes, $S_{\nu}$, of each feature, instead of the more usual brightness temperatures.
	
			\begin{table}[h]
			\small
			\centering
			\caption{Flux budget [SFU] for {\bf 1-Oct-2009} observations (sources are background-subtracted, relative errors are below the 10\% level):
			}
			\begin{tabular}{llll}
				\hline
				Frequency	& Full-Sun 			& Limb source in 	& Limb source		\\
						& (within 1.3 $R_S$)		& upper right quadrant 	& in lower--right quadrant\\
				\hline \hline
				2750 MHz	& 71.4 (100\%)			& 1.1 (1.6\%)		& 2.5 (3.5\%)		\\
				\hline
			\end{tabular}
			\label{tab:budget:Oct01}
			\end{table}

			\begin{table}[h]
			\caption{ Flux budget [SFU] for {\bf 29-Jan-2010} observations (sources are background-subtracted):
			}
			\begin{tabular}{lll}
				\hline
				Frequency	& Full-Sun flux		& Active Region flux\\
						& (within 1.3 $R_S$)	&		\\
				\hline \hline
				1430 MHz	& 57.5 (100\%)		& 3.8 (6.6\%)	\\
				2750 MHz	& 71.6 (100\%)		& 3.6 (5.0\%)	\\
				3500 MHz	& 89.5	(100\%)		& 3.6 (4.5\%)	\\
				\hline
			\end{tabular}
			\label{tab:budget:Jan29}
			\end{table}

			\begin{table}[h]
			\caption{Flux budget [SFU] for {\bf 21-Mar-2010} observations (sources are background-subtracted, see top left plot of Figure~\ref{fig:img:Mar21:4a} for letter labeling):
			}
			\centering
			\small
			\begin{tabular}{llllll}
				\hline
				Frequency	& Full-Sun		& Main AR	& Left limb	& Right limb, 		& Right limb, 		\\
						& 			&   ``A''	& ``B''		&  ``C''		& ``D''			\\
				\hline \hline				
				1430 MHz	& 61.3 (100\%)		& 5 (8.1\%)	& 2 (3.3\%)	& 1 (1.6\%)		& 0.5 (0.8\%)		\\
				2100 MHz	& 86.3 (100\%)		& 7.8 (9.0\%)	& 2.5 (2.9\%)	& 1.5 (1.7\%)		& 0.4 (0.5\%)	 	\\
				2750 MHz	& 76.3 (100\%)		& 6.0 (7.9\%)	& 1.7 (2.2\%)	& 1.1 (1.4\%)		& 0.35 (0.5\%)		\\
				3500 MHz	& 83.0 (100\%)		& 5.8 (7.0\%)	& 1.6 (1.9\%)	& 1.1 (1.3\%)		& 0.27 (0.3\%)		\\
				4499 MHz	& 114.7 (100\%)		& 5.2 (4.5\%)	& 1.7 (1.5\%)	& 1.6 (1.4\%)		& 0.8 (0.7\%)		\\
				6000 MHz	& 130.0 (100\%)		& 4.6 (3.5\%)	& 1.8 (1.4\%)	& 1.5 (1.2\%)		& 0.9 (0.7\%)		\\
				\hline
			\end{tabular}
			\label{tab:budget:Mar21}
			\end{table}

	Our observations show that for these particular cases, discrete background-subtracted ARs on the Sun typically contain a few percent of the total solar flux at various frequencies.
	The weaker ARs' generally flat spectra lead us to conclude that they were producing mostly optically thin thermal bremsstrahlung emission.
	However, the main AR in the 21-Mar-2010 observations (``A'' in Figure~\ref{fig:img:Mar21:4a}) possesses an additional component that exhibits an increase centered around 2.8 GHz, typical of GR emission.
	
	In the next section, we will merge these multi-frequency observations into a single quantity, that we dubbed ``Coronal Thermal Bremsstrahlung Contribution'' or ``$\eta$'' function,
	and use it to determine, amongst other things, the fraction of emission that is due to GR.

	\section{Coronal Thermal Bremsstrahlung Contribution ``CTBC''} \label{sect:CTBC}

		\subsection{Theory:}
		
		In the microwave range, the brightness temperature at Sun center can be modeled as the sum of an optically thick chromosphere ($\approx$10\,000 K)
		and an optically thin bremsstrahlung emission from an exponential hydrostatic corona ({\it e.g.} \opencite{Zirin1991}):
		\begin{equation}
			T_{\textrm{B}} = T_{\textrm{chromo}} + \tau \, T_{\textrm{corona}}
		\end{equation}
		with the free--free optical depth $\tau$ given by:
		\begin{equation}
			\tau = \zeta \, \nu^{-2} \, \frac{EM}{T^{3/2}}
		\end{equation}
		where $EM$ is the line-of-sight (LOS) emission measure, $\nu$ the frequency, $T$ the temperature of the coronal plasma, 
		and $\zeta=\zeta(T/\nu)$ a slowly varying function of $T/\nu$, approximately given by \cite{Dulk1985}:
		\begin{equation}
			\zeta(T/\nu) = 9.786 \times 10^{-3} \ln \left( 4.7 \times 10^{10} \times \frac{T}{\nu} \right)
		\end{equation}
		We have assumed the refractive index to be unity, appropriate for the microwave regime outside of active regions ({\it e.g.} \opencite{LandiChiuderiDrago2008} and \opencite{Shibasaki2011}).
		We define, for every pixel, the $CTBC$ ``$\eta$'' function as follows:
		\begin{equation}
			\eta(\nu) \equiv \frac{ T_{\textrm{B}}(\nu)-T_{\textrm{chromo}} }{\zeta \, \nu^{-2}}
		\end{equation}
		where $T_{\textrm{B}}(\nu)$ is the observed brightness temperature at frequency $\nu$, $T_{\textrm{chromo}}$ is set to 10\,880 K, the value fitted by \inlinecite{Zirin1991}, and, to compute $\zeta$, we assume a temperature of 1 MK.
		As stated earlier, $\zeta$ depends only weakly on the exact value of $T$.	

		For each pixel where optically thin free--free thermal emission is the only emission mechanism, one expects:
		\begin{equation}\label{eq:eta}
			\eta = \frac{EM}{\sqrt{T}}
		\end{equation}
		{\bf which is independent of frequency}.
		(More generally, $\eta$=$\Sigma \,\frac{EM_i}{\sqrt{T_i}}$ or \\ $\eta$ = $\int \frac{\left(\frac{\mathrm{d}EM}{\mathrm{d}T}\right)}{\sqrt{T}} \mathrm{d}T$).
		{\bf Provided the physical assumptions are valid, this does not depend on the spatial distribution of the coronal features.}
		In the absence of flares, deviations indicate other physical phenomena are at work: optical thickness (at low frequencies), 
		or gyroresonance emission \cite{Zlotnik1968a,Zlotnik1968b,WhiteKundu1997}.
		
		Furthermore, if $\eta(\nu)$ for a feature is indeed the same at all frequencies, 
		then the physics of the emission mechanism and the actual coronal density and temperature structure can be separated, 
		and the latter independently studied and modeled.

		\subsection{Observations:}

			\begin{figure}[p]
			\centering
			\includegraphics[width=10cm]{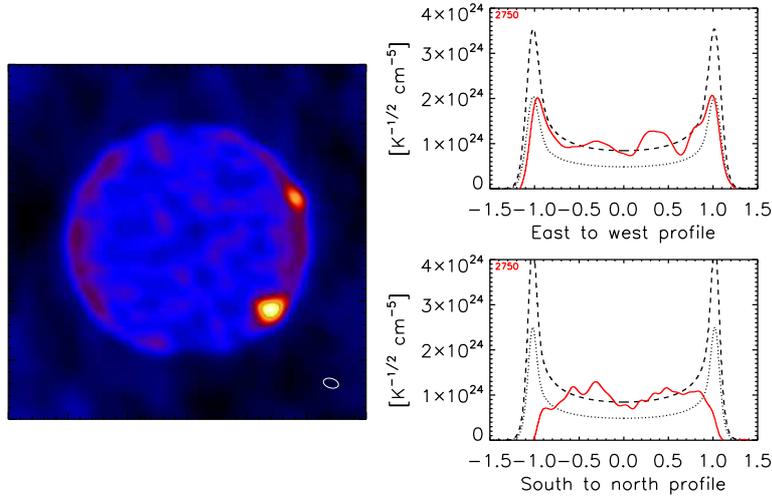}
			\caption{
				1-Oct-2009: $\eta$ maps (left) and profiles (right, top and bottom).
				Overplotted are expected $\eta$-profiles from a 1-MK exponential corona with base electron density 
				$n_0=4.4\times 10^8$ cm$^{-3}$ (dotted line) and $n_0=5.8\times 10^8$ cm$^{-3}$ (dashed line), convolved with ATA beamsize.
			}
			\label{fig:img:Oct01:4c}
			\end{figure}

			\begin{figure}[p]
			\centering
			\includegraphics[width=12cm]{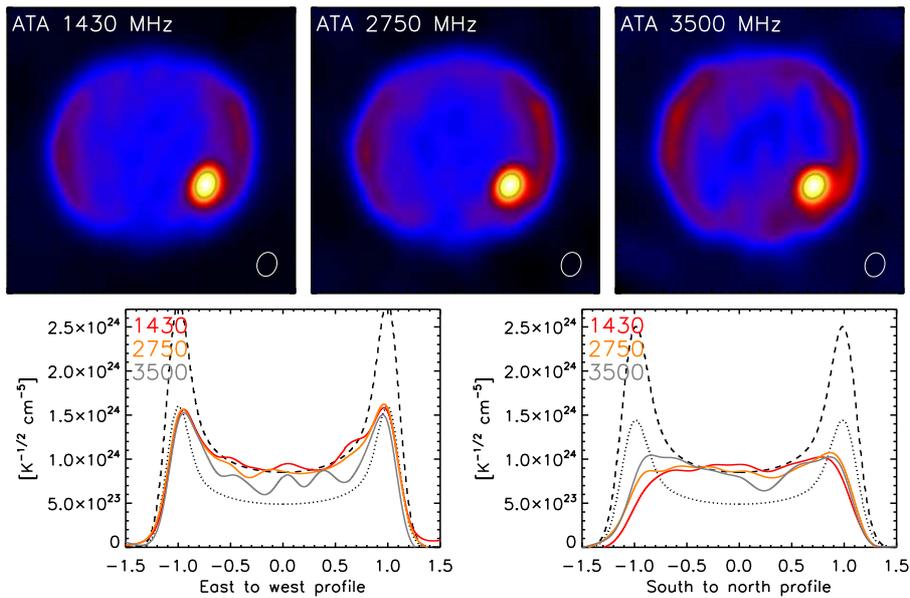}
			\caption{
				29-Jan-2010 $\eta$ maps and equatorial (lower left) and (lower middle) polar profiles.
				All maps (and hence their derived profiles also) were convolved to the largest beamsize (i.e. the 1430 MHz beam in this case).
				Overplotted are expected $\eta$-profiles from a 1-MK exponential corona with base electron density 
				$n_0=4.4\times 10^8$ cm$^{-3}$ (dotted line) and $n_0=5.8\times 10^8$ cm$^{-3}$ (dashed line), convolved with ATA beamsize.
			}
			\label{fig:img:Jan29:4c}
			\end{figure}
	
			\begin{figure}[h!]
			\centering
			\includegraphics[width=12cm]{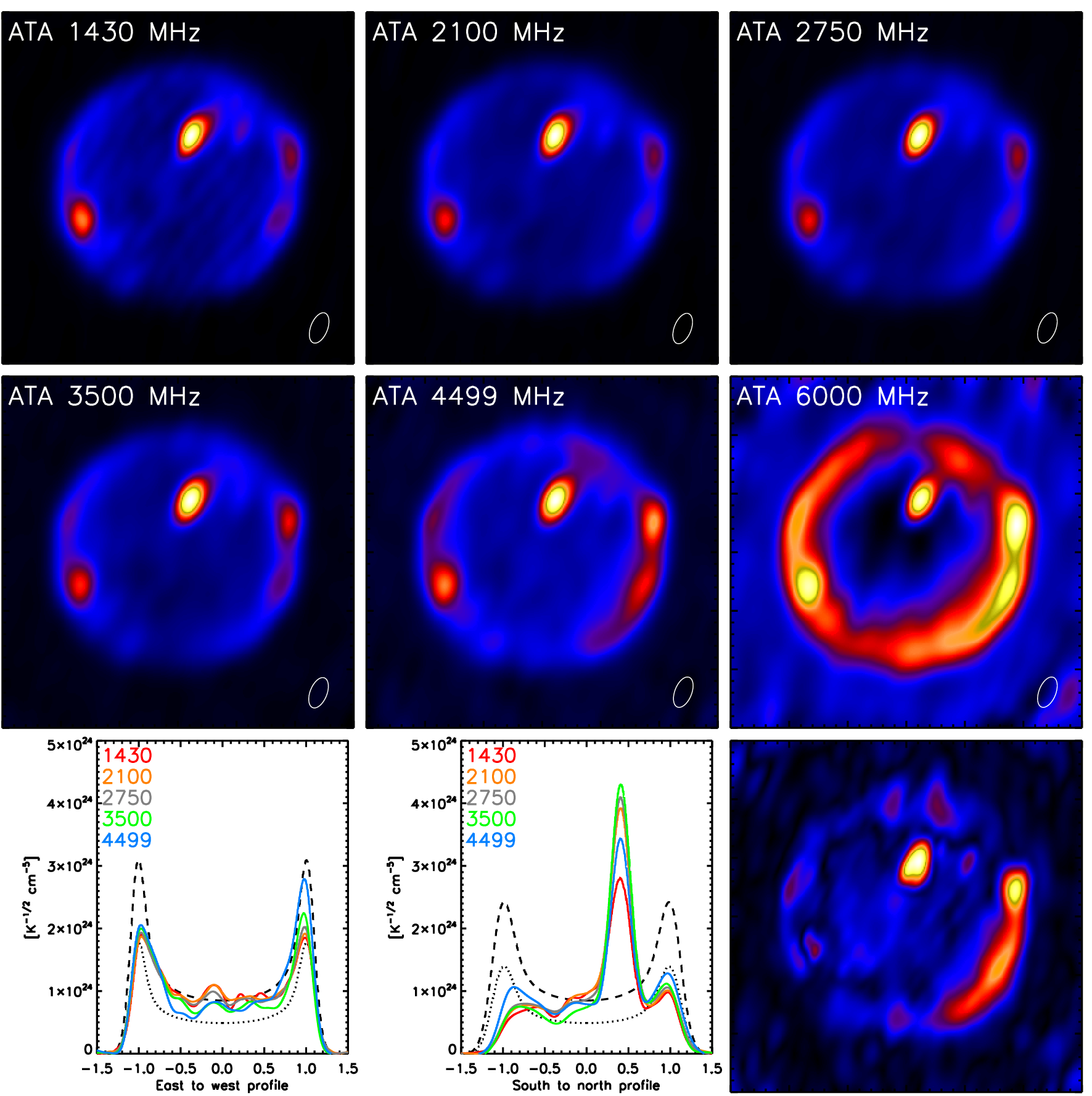}
			\caption{
				$\eta$ maps and profiles.
				All images were convolved to the same elliptical gaussian beam, 1.05 times the largest (1430 MHz) one.
				The bottom right image is the displays the standard deviation of the other six.
				Overplotted are expected $\eta$-profiles from a 1-MK exponential corona with base electron density 
				$n_0=4.4\times 10^8$ cm$^{-3}$ (dotted line) and $n_0=5.8\times 10^8$ cm$^{-3}$ (dashed line), convolved with ATA beamsize.
				Notice difference between equatorial and polar solid black profiles, due to beam anisotropy.
			}
			\label{fig:img:Mar21:4c}
			\end{figure}

		$\eta$-maps and profiles are displayed in Figures~\ref{fig:img:Oct01:4c}, \ref{fig:img:Jan29:4c}, and \ref{fig:img:Mar21:4c}, along with synthetic $\eta$-profiles for simple exponential coronae.
		For proper comparison, all maps were convolved to the worst point-spread-function (psf) of the set, i.e. to the psf of the lowest frequency.
		A first glance reveals that:
		\begin{itemize}
			\item There is rough agreement of the $\eta$-profiles across frequencies, consistent with Equation (\ref{eq:eta}). 
				The largest discrepancies appear in the vicinity of active regions (on-disk and at the limb) in the case of the 21-Mar-2010 observation.
			\item The $\eta$-value of the quiet Sun at disk center is about 8$\times$10$^{23}$ K$^{-1/2}$ cm$^{-5}$.
				The value of the limb spike is more complicated to estimate, as the profile is convolved with the instrument's psf, but can be estimated to be $\approx$10$^{25}$ K$^{-1/2}$ cm$^{-5}$. 
			\item The equatorial $\eta$-profiles can be constrained by hydrostatic exponential 1 MK coronae with base densities $\approx$4.4 and $\approx$5.8$\times$10$^8$ cm$^{-3}$,
				but neither faithfully reflect the actual profile.
			\item Equatorial limb brightening is observed, although some of it is due to the presence of active regions (21-Mar-2010 observations).
				At high frequencies, particularly 6 GHz, a ring appears. 
				As discussed earlier, the lack of short baselines at these frequencies makes the center of the Sun (large spatial structure) less bright than it should be, 
				increasing the contrast with the limbs, which are themselves better-imaged small scale structures.
		\end{itemize}

\begin{figure}[h!]
			\centering
			\includegraphics[width=12cm]{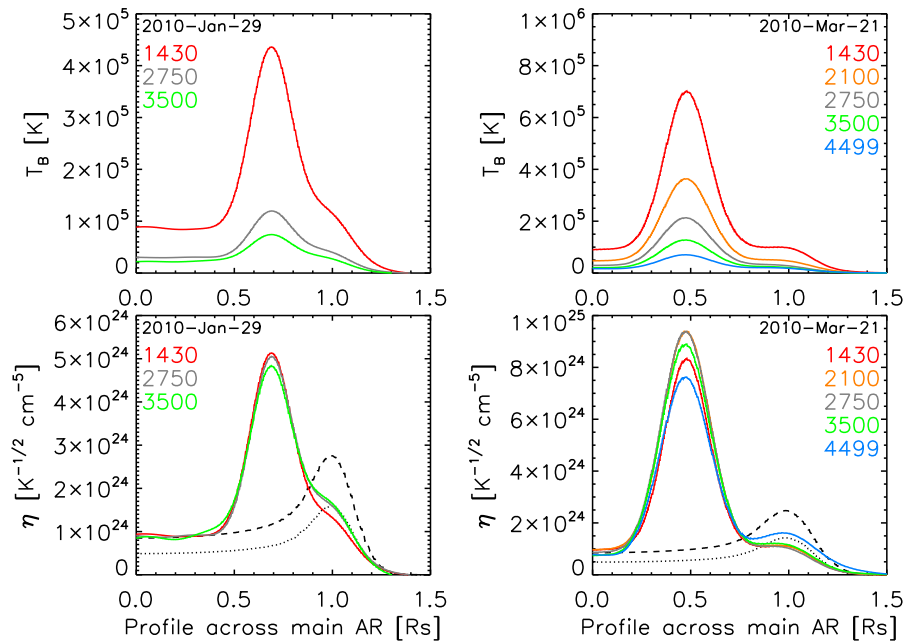}
			\caption{
				$T_{\textrm{B}}$ {\it (top row)} and $\eta$ {\it (bottom row)} radial profiles accross main AR for the 29-Jan-2010 {\it left column} and 21-Mar-2010 {\it right column} observations. 
				Otherwise, as in Figure~\ref{fig:img:Jan29:4c}.
			}
			\label{fig:prof:Jan29Mar21:mainAR}
			\end{figure}

			The $\eta$-profiles for 29-Jan-2010 appear to show little GR contribution (as they superpose well in all three frequencies, even in the AR -- Figure~\ref{fig:prof:Jan29Mar21:mainAR}). 
			From the AR's $\eta$-flux and its gaussian size $r_{\mathrm{source}}$ estimated from the observed size $r_{\mathrm{obs}}$ and the beamwidth $r_{\mathrm{beam}}$ using $r_{\mathrm{source}}=\sqrt{r_{\mathrm{obs}}^2-r_{\mathrm{beam}}^2}$, 
			the peak $\eta$ can be estimated, and amounts to $\approx$10$^{25}$ K$^{-1/2}$ cm$^{-5}$, which compares well with other observations (see Table~\ref{tab:losem}).
			Assuming a single temperature of $\approx$3 MK, this means an average line-of-sight emission measure (LOS EM) of $\approx$1.6$\times$10$^{28}$ cm$^{-5}$.
			
			\begin{table}[h]
			\caption{Approximate temperatures, emission measures, and $\eta$ values for various observations or models.
			}
			\centering
			\small
			\begin{tabular}{llccc}
				\hline
				Source			& Detail	& T 		& LOS EM 		& $\eta$ 				\\
							& 		& [MK]		& [cm$^{-5}$]		& [K$^{-1/2}$ cm$^{-5}$]		\\
				\hline \hline
				EUV observations,	& AR core	& 5		& 5$\times$10$^{28}$	& 2.2$\times$10$^{25}$			\\
				Aschwanden and Boerner 2011 & AR average	& 1.6		& 10$^{27}$		& 7.9$\times$10$^{23}$			\\
				
				\hline

				Exponential 1 MK	& Sun center	& 1		& 6.3$\times$10$^{26}$	& 6.3$\times$10$^{23}$			\\
				corona	with base density& limb spike	& 1		& 8.4$\times$10$^{27}$	& 8.4$\times$10$^{24}$			\\
				5$\times$10$^{8}$ cm$^{-3}$&		& 		&			&					\\ 
				
				\hline
				
				29-Jan-2010		& Main AR	& 		& 			& $\approx$10$^{25}$				\\
				ATA observations	&		& 		& 			& 					\\

				\hline

			\end{tabular}
			\label{tab:losem}
			\end{table}

			The spectrum of the 21-Mar-2010 disk AR (Figure~\ref{fig:sp:Mar21}, source ``A'': {\it red line}) indicates some gyroresonance contribution.
			The frequency-dependence of the $\eta$ maps over the disk AR (Figures~\ref{fig:img:Mar21:4c} and \ref{fig:prof:Jan29Mar21:mainAR}) and the presence of circular polarization (Figure~\ref{fig:img:Mar21:4a:V}) further support this.
			The $\eta$ value of the main AR is smallest at 6 GHz:
			this value can be be taken as upper boundary to the amount of FF emission in the disk AR.
			It is found that {\it at least} 63\% (representing 6.4\% of the total solar flux) of the 10.7 cm flux from the main AR is due to GR emission, the rest mostly from FF emission.
			If one assumes that there is {\it only} FF emission at 17 GHz, then an $\eta$-map derived from the Nobeyama Radioheliograph (NoRH, 17 GHz)
			at around 01:00 UT on 2010-Mar-22 gives us the correct FF flux, and hence the correct fraction of the emission that is due to GR: 76\% (or 7.8\% of total solar flux).

\section{Summary and Conclusions}\label{sect:ccl}

		The Allen Telescope Array has been used to obtain full--disk solar maps at several frequencies simultaneously.

		We have established a multi-frequency flux budget, finding that a single active region amounted for up to $\approx$8\% of the total solar flux at 2750 MHz, the actual fraction varying with frequency and active region.

		Using the multi-frequency capability, a simple method to find regions of the Sun where optically thin free--free emission dominates has been devised.
		Regions where the $\eta$-value varies with frequency are likely regions with a gyroresonance component.
		The emission from an active region observed on 29-Jan-2010 appeared to be purely free--free thermal bremsstrahlung in origin.
		However, the emission from an active region observed on 21-Mar-2010 appeared to be $\approx$75\% due to gyroresonance, and $\approx$25\% due to free--free emission at 2750 MHz.
		
		Although its current operational status is uncertain, the ATA's combination of high-quality mapping and frequency coverage put it in a unique position to address global coronal studies.
		Moreover, its high spectral resolution below 1 GHz enable it to explore a region that has been poorly investigated so far.


		

\begin{acks}
This work was supported by NASA Contract No. NAS 5-98033.
We thank the anonymous referee for comments that have helped to improve this article.
\end{acks}


\bibliographystyle{spr-mp-sola}

\end{article} 

\end{document}